\documentstyle[preprint,aps]{revtex}
\begin{document}

\bibliographystyle{unsrt}

\draft

\title{Ballistic Composite Fermions in Semiconductor Nanostructures}

\author{J.~E.~F.~Frost, C.-T.~Liang, D.~R.~Mace, M.~Y.~Simmons, 
D.~A.~Ritchie and M.~Pepper}
\address{Cavendish Laboratory,
Madingley Road, Cambridge CB3 0HE, United Kingdom}

\date{\today}

\maketitle

\begin{abstract}
We report the results of two fundamental transport measurements at a
Landau level filling factor $\nu$ of 1/2.  The well known ballistic
electron transport phenomena of quenching of the Hall effect in a
mesoscopic cross-junction and negative magnetoresistance of a
constriction are observed close to B~=~0 and $\nu~=~ 1/2$. The
experimental results demonstrate semi-classical charge transport by
composite fermions, which consist of electrons bound to an even number
of flux quanta.

\pacs{PACS numbers: 73.40.Hm, 73.40.Kp, 73.50.Jt}
\end{abstract}



Among the many experiments which demonstrate the ballistic nature of
electron transport in a clean two-dimensional sytem, reports by Ford
{\em et al\/} on quenching of the Hall effect in a
cross-junction\cite{quench3} and by van Houten {\em et al\/} on short
constriction negative magnetoresistance\cite{nmr1} are seminal. In
this Letter we report the first demonstration of these fundamental
effects at a Landau level filling factor $\nu$ of 1/2, where a
Chern-Simons gauge transformation maps the strongly interacting
electron system onto a system of weakly interacting composite fermions
in a zero effective magnetic field\cite{cft1,cft2}.

 In a small magnetic field applied perpendicular to the
two-dimensional electron gas (2DEG), magnetic focussing experiments
show that electrons travel with circular trajectories with a radius
$r_c~=~v_F/\omega_c$, where $v_F$ is the Fermi velocity and $\omega_c$
is the cyclotron frequency\cite{magfoc1}.  In a narrow wire, the Hall
voltage is found to ``quench'', rapidly dropping below its classical
value at some critical magnetic field\cite{quench1}. Control over the
precise geometry of a cross shaped sample can even result in a Hall
voltage of opposite sign to that predicted classically, an effect
attributed to a combination of ballistic electron transport and
largely specular reflection from the device edges channelling
electrons into the ``wrong'' voltage probe\cite{quench3}. In a
mesoscopic Hall bar, electron collimation and a small amount of
diffuse boundary scattering lead to a peak in the magnetoresistance at
an intermediate magnetic field before the onset of negative
magnetoresistance due to suppression of inter-edge
scattering\cite{boundary1}. A ballistic constriction exhibits negative
four-terminal magnetoresistance because with an increase in magnetic
field, a larger fraction of the edge states in the unpatterned 2DEG
are transmitted\cite{nmr1}.

At higher magnetic fields, the Hall resistance is found to take on
quantised values $R_H~=~h/$$\nu$$e^{2}$ with integral
$\nu$\cite{iqhe1}.  High mobility, low density 2DEG samples show a
rich structure in magnetoresistance with minima at $\nu$~=~p/q, with
integral p,q, and an increasing number of predominantly odd
denominator fractional minima have been resolved with their associated
plateaux in Hall resistance.  The odd denominator minima have been
explained in terms of a hierarchy of quasi-particle
states\cite{hierarchy1,hierarchy2}.

 At $\nu~=~1/2$, there is a broad minimum in magnetoresistance without
a corresponding Hall resistance plateau, and composite fermions are
thought to be the principal agents of charge transport. These
composite fermions, composed of electrons bound to an even number of
magnetic flux quanta, experience an effective zero magnetic field at
precisely $\nu~=~1/2$. \cite{cft1,cft2}.  Evidence continues to
accumulate for the existence of a Fermi surface at $\nu~=~1/2$ and
charge transport by composite
fermions\cite{cfe1,cfe2,cfe3,cfe4,cfe5,cfe6}.

 Experimental results were
obtained from split-gate type devices fabricated on two wafers grown
by MBE: T139 and A334.  Measurements were made after brief
illumination with a red LED in a pumped $^3He$ cryostat at 300 mK. A
current of 10nA and standard ac phase-sensitive detection was used for the
four-terminal resistance measurements.  Wafer T139 has sheet carrier
density $n_s = 1.3\times 10^{11}cm^{-2}$, mobility $\mu = 3.0\times
10^6cm^2V^{-1}s^{-1}$ and a 2DEG depth of 300 nm and wafer A334 has
$n_s = 1.2\times 10^{11}cm^{-2}$, $\mu = 1.8\times
10^6cm^2V^{-1}s^{-1}$ and a 2DEG depth of 300 nm.
 Figure {1(a)} shows the
geometry of the cross-junction, which consists of four symmetric
openings each 0.8$\mu$m wide. Figure {1(b)} shows the device used to
study a constriction, which is comprised of six split-gates in series
with a finger width of 0.3$\mu$m, pitch of 0.5$\mu$m and constriction
width of 1$\mu$m.  In the experiment, split-gates 2,4 and 6 are held
at a gate voltage of 0.6V, and split-gates 1,3 and 5 are held at a
negative gate voltage to give a voltage probe separation of 1$\mu$m.
The split-gates 2 and 4 above the voltage probes are held positive in
order to ensure that the sheet carrier density is not reduced
below that in the centre of the channel, eliminating unwanted
reflection of edge states in an applied magnetic field\cite{reflect1}.
The assumption that the voltage probes are ideal in this respect is
justified over the measurement range, as shown by the
presence of good zeroes in the magnetoresistance.


Figure {2} shows the Hall resistance $V_{CE}/I_{SD}$ with zero applied
gate voltage for the cross geometry sample.  The Hall resistance is
linear both in the vicinity of B~=~0 and $\nu~=~1/2$, with well
developed quantised Hall plateaux away from these magnetic fields. Using the
 sheet resistivity $\rho_{xx}$ at $\nu~=~1/2$ ($400~\Omega$/square) we estimate the composite
 fermion mean free to be approximately $1~\mu$m, larger  than the distance
 across the junction. The
insets of figure {2} show the Hall resistance and numerical derivative
$dR_{xy}/dB$ (a) near $B~=~0$ and (b) near $\nu~=~1/2$ when the cross-
junction is defined. The magnetic field scale of inset (b) is reduced
by a factor $\sqrt{2}$ to account for the spin polarised enhancement
in Fermi wave-vector at $\nu~=~1/2$\cite{cft2,cft3}.  Lithographic
imperfections lead to a slight asymmetry in the Hall resistance about
$B~=~0$ and $\nu~=~1/2$ and so gate biasses are adjusted to
compensate, but the same gate voltages are used at $B~=~0$ and at
$\nu~=~1/2$. With applied gate voltages of $V_{1}~=~V_{2}~=~-1.7~V$
and $V_{3}~=~V_{4}~=~-1.5~V$, a cross shaped junction is defined and
quenching of the Hall effect is observed at $B~=~0$. Quenching close
to $\nu~=~1/2$ is not so strong, but the deviation from linearity is
shown qualitatively by a minimum in $dR_{xy}/dB$ at $\nu~=~1/2$,
demonstrating the ballistic nature of charge transport. When an
electric current flows in a composite fermion system, an induced
effective electric field arises from the current of magnetic flux
quanta pairs\cite{cft3}.  The Hall resistance therefore remains finite
at $\nu~=~1/2$, with a value of $2h/e^{2}$ and we only observed a
minimum in the Hall {\em slope} at $\nu~=~1/2$, compared to the zero
in Hall {\em resistance} at $B~=~0$.


We now discuss the results of measurements of the ballistic
constriction. Figure {3} shows the longitudinal magnetoresistance
$V_{AB}/I_{SD}$ of the constriction device with
$V_2~=~V_4~=~V_6~=~0.6V$ for fixed voltages of $V_1~=~V_3~=~V_5$ = $-0.2$V, $-1.0$, $-1.4$, $-1.8$, $-2.6$V and $-3$V.
 Measurements using voltage probes E and F were similar to those using probes A and B. When
the 2DEG is depleted beneath gates 1,3 and 5, the device resembles a
mesoscopic Hall bar with a width and length of 1$\mu$m and the present
results may be compared with those in the literature for such
structures\cite{boundary1,boundary2}.  The small size of the active
region also minimizes unwanted effects due to wafer non-uniformity. A
shift of the $\nu~=~1$ Shubnikov-de Haas zero (at about 5T) towards a
lower magnetic field indicates that the sheet carrier density in the
channel drops from $1.2 $ to $ 1.1 \times 10^{11}cm^{-2}$ over this
range of gate voltage. The magnetoresistance at the smallest negative
defining gate voltage resembles that of a macroscopic Hall bar with a
shallow minimum in magnetoresistance at $B~=~0$ and $\nu~=~1/2$.

Macroscopic 2DEG samples typically show a longitudinal resistivity two
orders of magnitude greater at $\nu~=~1/2$ than at $B~=~0$\cite{cfe3}.
Random fluctuations in carrier concentration causing a corresponding
fluctuation in effective magnetic field and an increase in effective
mass at $\nu~=~1/2$ both contribute to an enhanced scattering rate for
composite fermions. We believe that this is the reason for the
presence of a single broad peak at $\nu~=~1/2$, compared with the
double peaks at $B~=~0$\cite{boundary2}. There is also a peak at
$\nu~=~3/2$, but it is less well defined than at $\nu~=~1/2$, in a
similar fashion to the observation of weak commensuribility
oscillations at $\nu~=~3/2$ by Kang {\em et al\/}\cite{cfe3}.
Deleterious effects due to the high series resistance of the
unpatterned 2DEG at $\nu~=~1/2$ are minimised by the use of voltage
probes in close proximity to the constriction under
investigation\cite{rfqhe1}.

As the defining gate voltage is made more negative, the double peak
structure develops close to B~=~0 and broad single peaks develop at
$\nu~=~1/2$ and $\nu~=~3/2$, indicated by circles and triangles
respectively in figure {3}. Landauer-B\"{u}ttiker formalism states
that scattering of electrons is necessary to establish local
equilibrium between voltage probe and sample\cite{LB1,LB2}. 
A difference in chemicalpotential between voltage probes
is not established in a mesoscopic Hall bar at $B~=~0$ if
collimation of ballistic electrons occurs, and a longitudinal
four-terminal resistance minimum results. The double peak structure
close to B~=~0 has been observed before only in the highest mobility
mesoscopic Hall bars where a non-specular component of the boundary
scattering gives a peak in resistance when $W/r_c~=~0.55$, where $W$
is the effective Hall bar width\cite{boundary1,boundary2}. In the
present work, the measured peak value at $\pm0.050T$ implies an
effective channel width of $0.6\mu$m, comparable with the lithographic
dimension. The magnetoresistance structure is symmetric about $B~=~0$
and this symmetry remains about $\nu~=~1/2$ and $\nu~=~3/2$,
particularly for the largest gate voltages and supports recent theory
predicting effective magnetic fields of opposite sign about
$\nu~=~1/2$\cite{cft2}. We suggest that composite fermion negative effective
magnetic field effects are only observed when a semi-classical
trajectory does not cross a boundary between positive and negative
effective field between voltage probes\cite{rfqhe1}.

The magnetoresistance for $\mid~B~\mid~<~0.4T$ excluding the central
minimum is well described by the equation giving the four terminal
resistance of a ballistic constriction, \begin{equation}
R_{4t}~=~(h/e^2)(1/N_{min}-1/N_{max}), \end{equation} where $N_{min}$
and $N_{max}$ are the number of occupied one-dimensional(1D) subbands in
the channel and unpatterned 2DEG respectively\cite{nmr1}. The
enhancement of composite fermion scattering results in broader single
peaks at high magnetic field. The peak heights at $B~=~0$, $\nu~=~1/2$
and $\nu~=~3/2$ increase with an increase in gate voltage as the
number of occupied 1D subbands in the constriction decreases,
according to equation {1}. These results are consistent with ballistic
composite fermion transport and the formation of 1D composite fermion
subbands in a constriction at $\nu~=~1/2$.

 Low-temperature four-terminal
magnetoresistance measurements have been performed on a mesoscopic
cross-junction and a constriction defined by Schottky gate
metallisation above a two-dimensional electron gas in a GaAs/AlGaAs
heterostructure with low sheet carrier density and high mobility.  The
longitudinal and Hall resistance at small applied magnetic field are
compared with that close to Landau level filling factors of $\nu$~=~1/2
and $\nu$~=~3/2 as the structures are defined.

For the cross geometry sample, the onset of quenching is observed as
nonlinearity in the Hall resistance both at low B and near a Landau
level filling factor of 1/2. For the constriction, peaks in the
longitudinal magnetoresistance are observed both near $B~=~0$ and when
$\nu~=~1/2$.  The effects are more pronounced as the
confining gate voltage is increased in magnitude.

Both phenomena occur due to the influence of sample geometry on the
semi-classical ballistic charge carrier trajectories.  We propose that
analogous mechanisms apply both near $B~=~0$ and near $\nu~=~1/2$ and
$\nu~=~3/2$, with composite fermions as the principal agents of charge
transport at high magnetic field rather than electrons.

This work was funded by the United
Kingdom(UK) Engineering and Physical Sciences Research Council.  We would
like to thank C.H.W. Barnes, I.M. Castleton, C.J.B. Ford, B.I.
Halperin, G. Kirczenow, J.T. Nicholls, C.G. Smith and V.I. Talyanskii
for useful discussions, A.R. Hamilton and B. Kardynal for experimental
assistance and D. Heftel, A. Beckett and D.R. Charge for technical
support. C.T.L acknowledges support from Hughes Hall and the committee
of the Vice-Chancellors and Principals, UK


\centerline{\bf Figure Captions}

Figure 1

Schematic diagrams of (a) the cross-junction device, (b) the
constriction device.

Figure 2

Hall resistance of the cross-junction device
with zero applied gate voltage.  Inset: Hall resistance (left scale)
and $dR_{xy}/dB$ (right scale) with
applied gate voltages of $V_{1}~=~V_{2}~=~-1.7~V$ and
$V_{3}~=~V_{4}~=~-1.5~V$, (a) around $B~=~0$ and (b) near
$\nu~=~1/2$.

Figure 3

Longitudinal magnetoresistance of the constriction device
with $V_2~=~V_4~=~V_6~=~0.6V$ for fixed voltages of $V_1~=~V_3~=~V_5$
= $-0.2$V, $-1.0$V, $-1.4$V, $-1.8$V, $-2.6$V and $-3$V from lowest to uppermost
trace respectively. Curves are offset by 300$\Omega$
 for clarity.


\begin{references}

 \bibitem{quench3}	C. J. B.  Ford, S.  Washburn, M.  B\"{u}ttiker,
C. M. Knoedler, and J. M. Hong, \prl {\bf 62}, 2724(1989). 

 \bibitem{nmr1} 	H. van Houten, C. W. J. Beenakker, P. H. M.
Loosdrecht, T. J. Thornton, H. Ahmed, M.  Pepper, C. T. Foxon and J. J.
Harris, \prb {\bf 37}, 8534 (1989). 

 \bibitem{cft1}		J. K. Jain, \prl {\bf 63}, 199 (1989).

 \bibitem{cft2}		B. I.Halperin, P. A. Lee and N. Read, Phys.Rev.
B {\bf 47}, 7312 (1993).

 \bibitem{magfoc1} 	 H. van Houten, C. W. J. Beenakker, J. G.
Williamson, M. E. I. Broekaart, P. H. M. Loosdrecht, B. J. van Wees, J. E.
Mooij, C. T. Foxon, and J. J. Harris, Europhys. Lett. {\bf 5}, 721
(1988). 

 \bibitem{quench1}	M. L. Roukes, A. Scherer, S. J. Allen Jr., H. G.
Craighead, R. M. Ruthen, E. D. Beebe, and J. P. Harbison, \prl {\bf 59},
3011 (1987).

 \bibitem{boundary1}	T.  J. Thornton, M. L. Roukes, A. Scherer, and
B. P. Van der Gaag, \prl {\bf 63}, 2128 (1989).  

\bibitem{iqhe1}		K. von Klitzing, G. Dorda, and M. Pepper,
\prl {\bf 45}, 494 (1980).

\bibitem{hierarchy1}	F. D. M. Haldane, \prl {\bf 51}, 605 (1983).

 \bibitem{hierarchy2}	B. I. Halperin, \prl {\bf 52}, 1583 (1984).

 \bibitem{cfe1}		R. R. Du, H. L. St\"{o}rmer, D. C. Tsui, L. N.
Pfeiffer, and K. W. West, \prl {\bf 70}, 2944 (1993).

 \bibitem{cfe2}		R. L. Willett, R. R. Ruel, K. W. West, and L. N.
Pfeiffer, \prl {\bf 71}, 3846 (1993).

 \bibitem{cfe3}		W. Kang, H. L. St\"{o}rmer, L. N. Pfeiffer, K. W.
Baldwin, and K. W. West, \prl {\bf 71}, 3850 (1993).
  
 \bibitem{cfe4}		D. R.Leadley, R. J. Nicholas, C. T. Foxon, and
J. J. Harris, \prl {\bf 72}, 1906 (1994).

 \bibitem{cfe5}		R. R. Du, H. L. St\"{o}rmer, D. C. Tsui, L. N.
Pfeiffer, and K. W. West, Solid State Commun. 90, {\bf 71} (1994).

 \bibitem{cfe6}		V. J. Goldman, B. Su, and J. K. Jain, \prl {\bf
72}, 2065 (1994).

 \bibitem{reflect1} B. J. van Wees, E. M.  M. Willems, C. J. P. M. Harmans,
C. W. J.Beenakker, H. van Houten, J. G. Williamson, C. T. Foxon, and J. J.
Harris, \prl {\bf 62}, 1181 (1989).

\bibitem{cft3} G. Kirczenow, and B. L. Johnson, \prb {\bf 51}, 17579 (1995).

\bibitem{boundary2} 	J. A. Simmons, D. C. Tsui, and G. Weimann, 
Surf. Sci. {\bf 196}, 81 (1988). 

\bibitem{rfqhe1}	J. E. F. Frost, C.-T. Liang, D. R. Mace, M. Y. Simmons,
D. A. Ritchie, and M. Pepper, Solid State Comm. {\bf 96}, 327 (1995).

\bibitem{LB1}		R. Landauer, IBM J. Res. Dev. {\bf 1}, 223 (1957).

\bibitem{LB2}		M. B\"{u}ttiker, \prl {\bf 57}, 1761 (1986).
 \end{references}
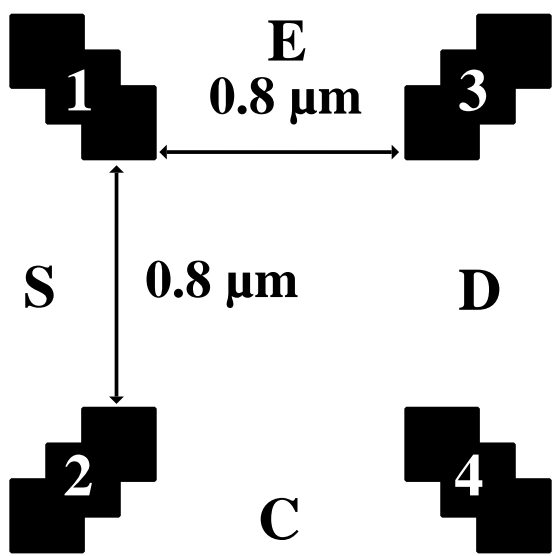
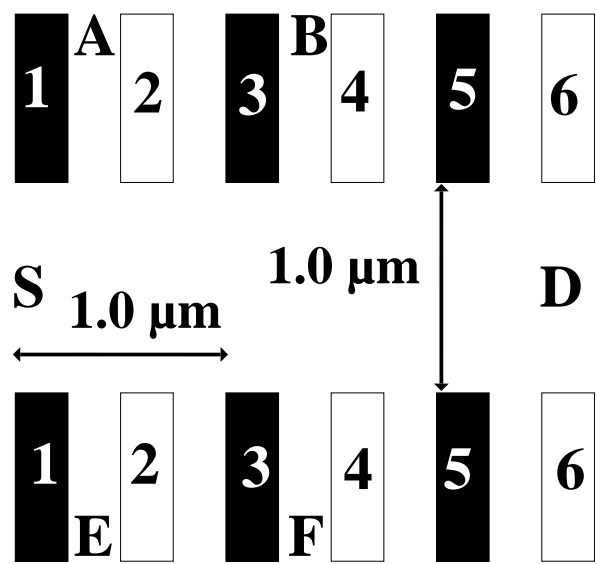
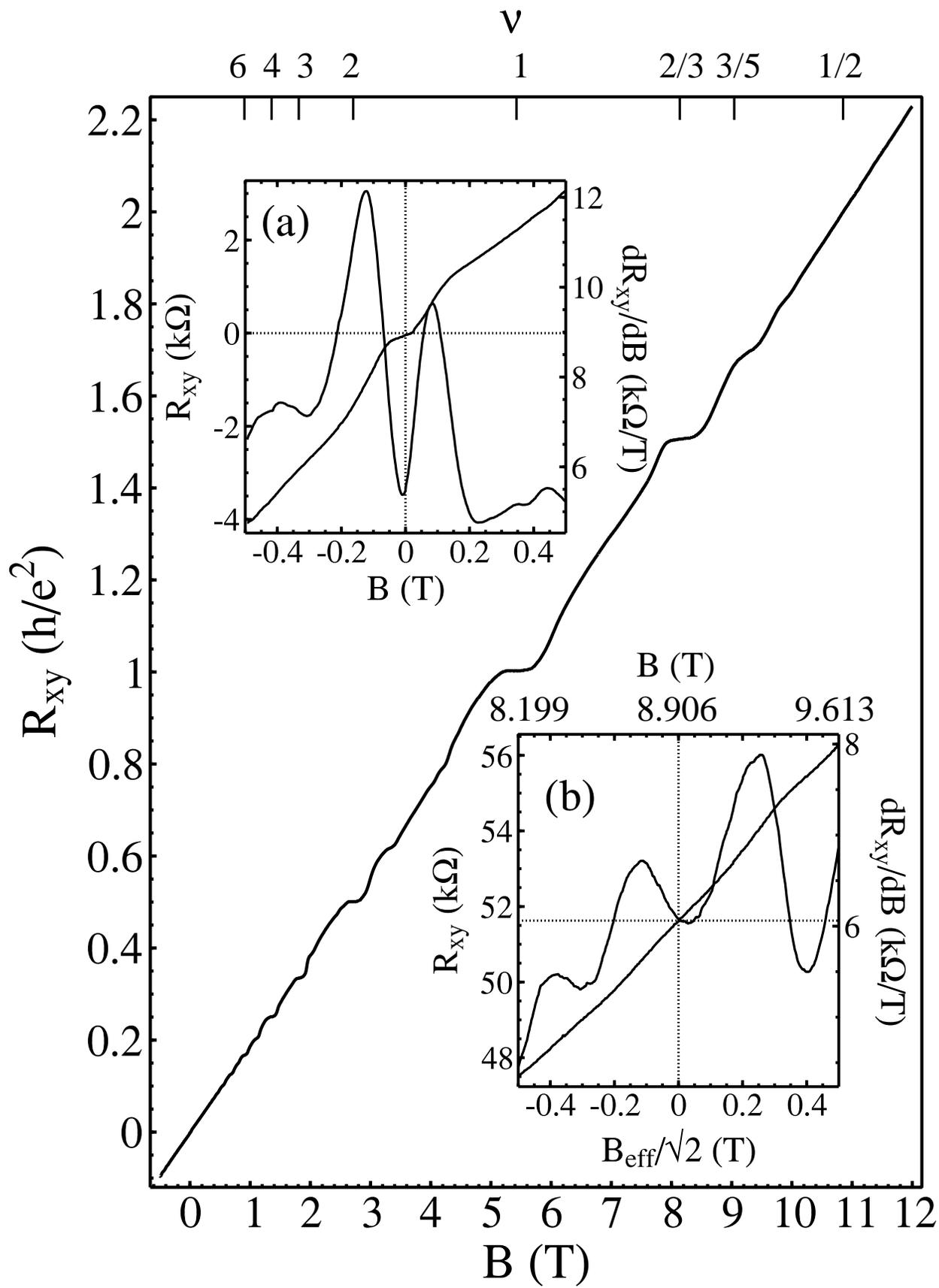
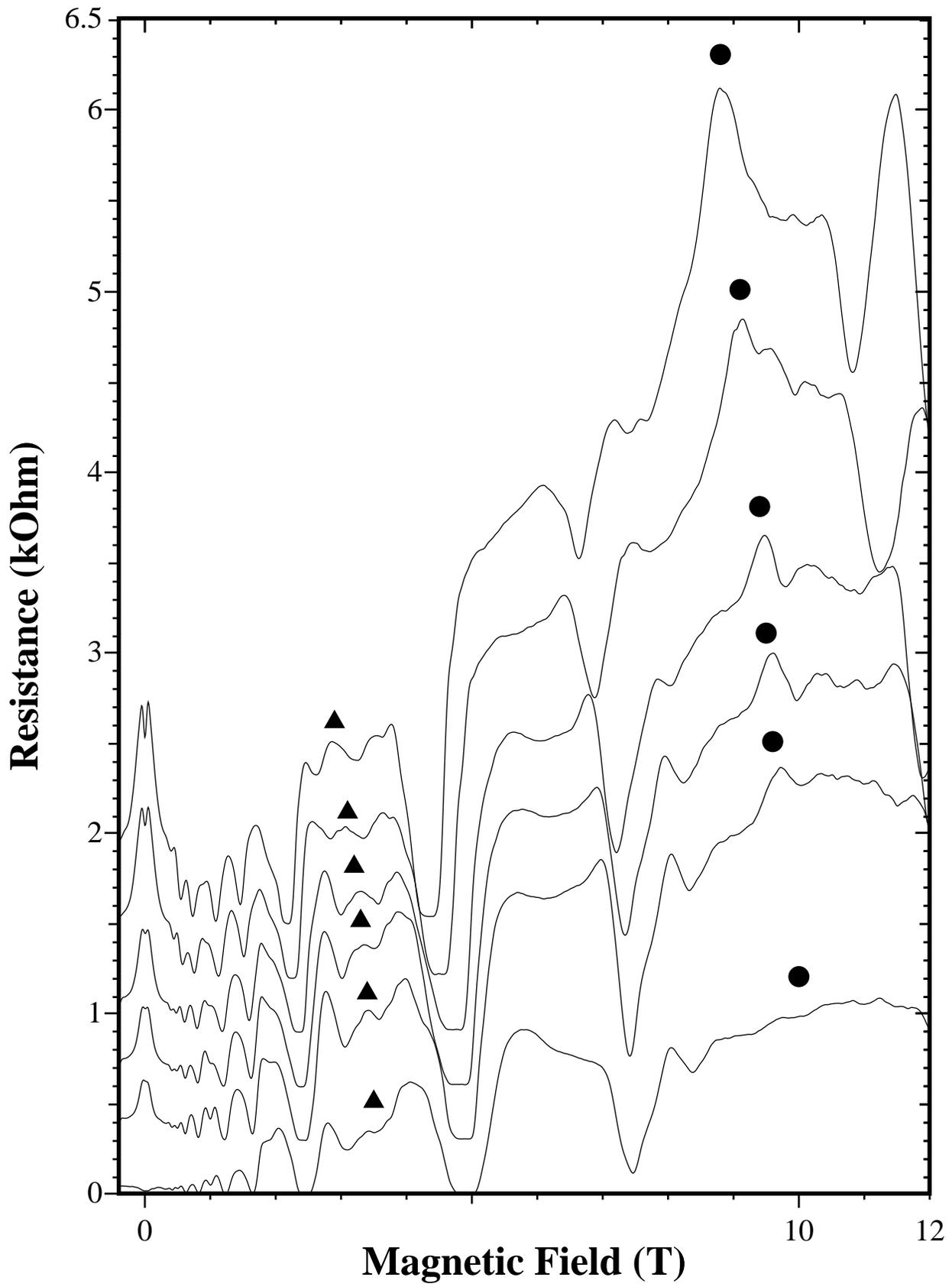
\end{document}